\documentclass[sigconf]{acmart}
\usepackage{booktabs} 
\usepackage{enumitem}
\usepackage{makecell}
\usepackage{textcomp}
\usepackage{amsmath}
\usepackage{algorithm} 
\usepackage[noend]{algpseudocode}
\usepackage{pdfpages}
\usepackage{graphicx}
\usepackage{mathrsfs}
\usepackage{xcolor}
\usepackage{epsfig}
\usepackage{graphicx}
\usepackage{multirow}

\usepackage{enumerate}
\usepackage{flushend}

\usepackage[normalem]{ulem}
\usepackage{subfigure}

\usepackage{lineno}

\AtBeginDocument{%
  \providecommand\BibTeX{{%
    \normalfont B\kern-0.5em{\scshape i\kern-0.25em b}\kern-0.8em\TeX}}}

\copyrightyear{2024}
\acmYear{2024}
\setcopyright{acmlicensed}\acmConference[CIKM '24]{Proceedings of the 33rd ACM International Conference on Information and Knowledge Management}{October 21--25, 2024}{Boise, ID, USA}
\acmBooktitle{Proceedings of the 33rd ACM International Conference on Information and Knowledge Management (CIKM '24), October 21--25, 2024, Boise, ID, USA}
\acmDOI{10.1145/3627673.3679792}
\acmISBN{979-8-4007-0436-9/24/10}

\begin{document}

\title{Reformulating Conversational Recommender Systems as Tri-Phase Offline Policy Learning}

\settopmatter{authorsperrow=3}

\author{Gangyi Zhang}
\affiliation{%
 \institution{University of Science and Technology of China}
 \city{Hefei}
 \country{China}
 }
\email{gangyi.zhang@outlook.com}

\author{Chongming Gao}
\authornote{Corresponding author.}
\affiliation{%
 \institution{University of Science and Technology of China}
 \city{Hefei}
 \country{China}
 }
\email{chongminggao@ustc.edu.cn}

\author{Hang Pan}
\affiliation{%
 \institution{University of Science and Technology of China}
 \city{Hefei}
 \country{China}
 }
\email{hungpaan@mail.ustc.edu.cn	}

\author{Runzhe Teng}
\affiliation{%
 \institution{University of Science and Technology of China}
 \city{Hefei}
 \country{China}
 }
\email{trz177@mail.ustc.edu.cn}

\author{Ruizhe Li}
\affiliation{%
 \institution{University of Science and Technology of China}
 \city{Hefei}
 \country{China}
 }
\email{imlrz@mail.ustc.edu.cn}





\begin{abstract}

Existing Conversational Recommender Systems (CRS) predominantly utilize user simulators for training and evaluating recommendation policies. These simulators often oversimplify the complexity of user interactions by focusing solely on static item attributes, neglecting the rich, evolving preferences that characterize real-world user behavior. This limitation frequently leads to models that perform well in simulated environments but falter in actual deployment. Addressing these challenges, this paper introduces the Tri-Phase Offline Policy Learning-based Conversational Recommender System (TCRS), which significantly reduces dependency on real-time interactions and mitigates overfitting issues prevalent in traditional approaches. TCRS integrates a model-based offline learning strategy with a controllable user simulation that dynamically aligns with both personalized and evolving user preferences. Through comprehensive experiments, TCRS demonstrates enhanced robustness, adaptability, and accuracy in recommendations, outperforming traditional CRS models in diverse user scenarios. This approach not only provides a more realistic evaluation environment but also facilitates a deeper understanding of user behavior dynamics, thereby refining the recommendation process.

\end{abstract}

\maketitle

\section{Introduction}

\label{sec:introduction}
\begin{figure*}[t]
    \centering
	\includegraphics[width=\linewidth]{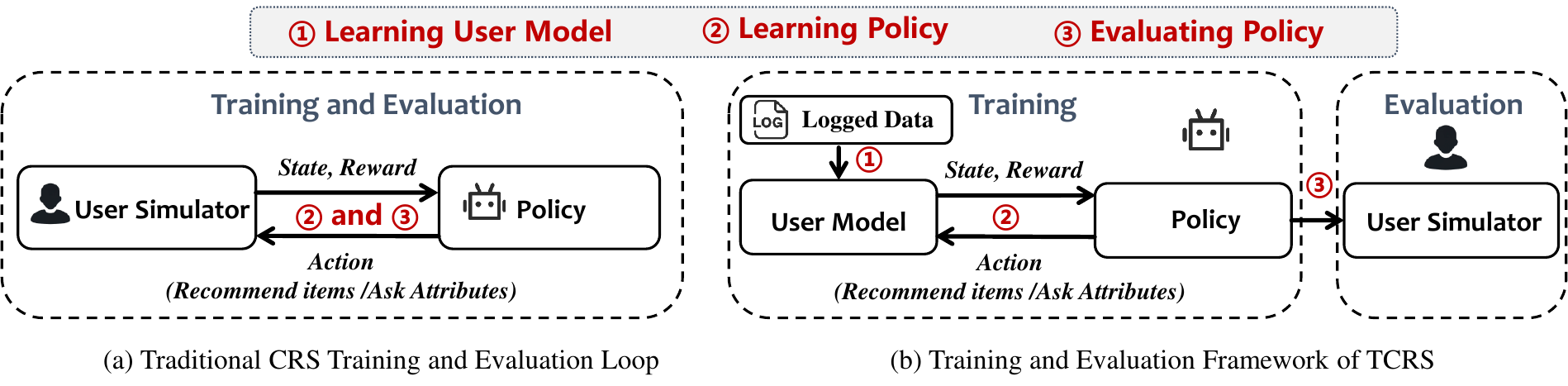}
    \caption{Comparative Frameworks in CRS Training and Evaluation}
    \label{fig:intro-exa}
    \vspace{-10pt}
\end{figure*}

Conversational Recommender Systems (CRS) have revolutionized traditional recommendation systems by enabling dynamic and interactive conversations with users to elicit fine-grained preferences and provide more accurate and explainable recommendations \cite{gao2021advances}. The effectiveness of CRS hinges critically on the availability and quality of real-time user interaction data \cite{gao2022kuairec, christakopoulou2016towards}, which poses significant challenges—data collection is often expensive and intrusive, potentially degrading the user experience. Traditionally, as illustrated in Figure \ref{fig:intro-exa}(a), CRS models have relied heavily on user simulators during both training and evaluation phases to mimic real-world interactions.
However, this reliance on the exact same simulator for both training and evaluation is problematic. It results in the leakage of evaluation rules and information to the policy during the training phase. Ideally, policy learning should be conducted offline, where the policy cannot interact with or explore the environment or users during training \cite{gao2023alleviating}. Instead, it should rely solely on logged data for learning.
Furthermore, existing simulators predominantly focus on static user preferences and fail to account for the evolving and individualized nature of user preferences. Consequently, systems trained and evaluated using such simulators perform admirably in controlled, simulated environments but often falter when deployed in real-world settings where user preferences are neither static nor predictable \cite{gao2023cirs}.

Recognizing these challenges, we reformulate CRS as a Tri-Phase Offline Policy Learning-based Conversational Recommender System (TCRS). It is a novel CRS architecture designed to drastically reduce reliance on real-time user interactions and enhance the robustness and accuracy of CRS models. TCRS integrates a model-based offline policy learning strategy with a controllable user simulation that dynamically adapts to both historical and evolving user preferences. This approach not only addresses the critical flaws of overfitting and unrealistic preference modeling inherent in traditional CRS but also provides a more realistic evaluation environment through its advanced simulation techniques.

As depicted in Figure \ref{fig:intro-exa}(b), the TCRS framework consists of three key components: 1) \textit{User Model Training}, which estimates user preferences from offline data and captures the dynamic and personalized nature of user preferences; 2) \textit{Policy Learning}, which employs the learned user model as a simulated environment to facilitate the training of recommendation policies; and 3) \textit{Policy Evaluation}, which assesses the trained policies using an independent, black-box simulator that dynamically adjusts to user interactions for more realistic evaluation.
By decoupling the training and evaluation phases, TCRS aims to mitigate the overfitting issue associated with traditional simulator-based approaches. Additionally, the explicit modeling of user preference dynamics and personalization within the user model can help CRS policies better capture real-world user behavior, leading to improved recommendation performance when deployed in actual user environments.

Building upon the overall TCRS framework, we now describe the specific implementation details of each key component:

1. \textbf{Conversational User Model}: This component employs a multi-task learning approach to capture and predict user preferences from a rich dataset of offline interactions, comprising a State Tracker for maintaining a comprehensive representation of user preference state, and a Preference Ranking for jointly optimizing item-level and attribute-level preference prediction.

2. \textbf{Conversational Policy Learning}:  We implement a Proximal Policy Optimization (PPO) algorithm to derive effective decision-making strategies, where the policy takes the preference state and preference-based reward as input to learn actions (recommended items or queried attributes) that maximize long-term reward.

3. \textbf{Controlable User Simulation}: We design a rule-based, user-centric preference simulator that dynamically aligns with the dual-perspective approach of historical personal preferences and target items preferences. This simulator, using personalization preference parameter and preference evolution rate, provides a critical black-box testing environment to evaluate the model's adaptability across various user scenarios.

Our key contributions are as follows:
\begin{itemize}[leftmargin=*]
\item  \textbf{A Paradigm Shift in CRS Training and Evaluation}: By decoupling the training and evaluation phases, TCRS introduces a novel offline policy learning framework that is both efficient and scalable, addressing the critical flaws of overfitting and unrealistic preference modeling inherent in traditional CRS approaches.
\item \textbf{More Realistic Evaluation Environment}: The Controllable User Simulation module in TCRS transcends the static, item-centric assumptions of previous simulators, embracing the personalized and evolving nature of user preferences during interactions. This allows for a comprehensive assessment of policy performance in diverse and realistic user scenarios.
\item \textbf{Versatile Policy Evaluation}: The Controllable User Simulation enables the creation of varied user preference environments, allowing us to thoroughly evaluate the adaptability and robustness of our CRS policies to different personalized and evolving user preferences. 
\end{itemize}

These innovations pave the way for more effective and realistic CRS that can better understand and adapt to individual user needs, ultimately leading to enhanced user experiences and increased adoption of CRS in real-world settings.

\section{Related Work}\label{sec:related}

\textbf{Conversational Recommender Systems (CRS)} aim to dynamically interact with users to elicit their preferences and provide personalized recommendations \cite{gao2021advances}. The research in CRS can be broadly divided into two approaches: policy optimization-based methods \cite{christakopoulou2018q, Sun:2018:CRS:3209978.3210002,lei20estimation,lei2020interactive,deng2021unified,zhang2022multiple,zhang2023adaptive, li2020seamlessly, chu2023meta, yang2024generate} and dialogue generation-based methods \cite{nips18/DeepConv, chen-etal-2019-towards, zhou2020improving,liao2020topic}. Policy Optimization-based Methods emphasize real-time, dynamic capturing of user preferences through strategic interactions \cite{lei2020interactive,wu2022state}. 
Reinforcement learning techniques \cite{lei2020interactive, deng2021unified} are commonly employed to develop effective recommendation policies. These methods focus on maximizing long-term user satisfaction by learning to ask purposeful questions that reveal user preferences effectively \cite{christakopoulou2018q, Sun:2018:CRS:3209978.3210002}. Dialogue Generation-based Methods use natural language processing (NLP) techniques to generate fluent and contextually appropriate dialogues \cite{nips18/DeepConv, chen-etal-2019-towards, zhou2020improving, dao2024broadening}. While these methods can create natural interactions, they often struggle with decision-making capabilities \cite{friedman2023leveraging}. The reliance on large dialogue corpora limits their ability to handle specific user contexts and detailed preference modeling \cite{nips18/DeepConv}.

Our research prioritizes policy optimization-based CRS due to its superior ability to adapt to user preferences dynamically and make informed decisions during interactions \cite{gao2021advances}. This aligns with our goal of enhancing the effectiveness and realism of CRS through robust offline policy learning.

\textbf{Offline Reinforcement Learning (Offline RL)} has gained traction in the recommendation system domain due to its ability to leverage historical data for policy training without the need for continuous online interactions \cite{gao2023cirs}. This approach addresses the challenges of data scarcity and user experience degradation associated with real-time data collection. Various methods integrate offline RL with recommendation systems, focusing on balancing exploration and exploitation \cite{gao2023alleviating}, handling cold-start problems \cite{chu2023meta}, and improving recommendation accuracy through advanced policy learning techniques \cite{deng2021unified, xin2022rethinking}. These methods often rely on historical interaction data to train robust policies capable of adapting to new user preferences and scenarios \cite{deffayet2023offline}.

Our TCRS framework builds upon these offline RL methodologies by introducing a tri-phase policy learning approach for CRS. This approach decouples the training and evaluation phases, leveraging a model-based user simulator to enhance policy robustness and accuracy without the dependency on real-time user data \cite{EasyRL4Rec, billp2024}.

\textbf{User Simulators} play a critical role in training and evaluating CRS policies by providing synthetic user feedback \cite{lei20estimation}. The design and realism of these simulators significantly impact the effectiveness of the trained policies \cite{gao2023cirs, gao2022kuairec}. Traditional CRS user simulators often assume static user preferences tied directly to item attributes \cite{lei2020interactive, zhang2022multiple}. These simulators generate feedback based solely on predefined rules, which can lead to overly optimized policies that perform well in simulations but fail to generalize to real-world interactions. Recent advancements advocate for more realistic simulators \cite{zhang2023adaptive, zhang2023user, deng2024towards} that capture the dynamic and personalized nature of user preferences. These simulators model user behavior as evolving and context-dependent, providing a more comprehensive evaluation environment \cite{zhang2020evaluating} for CRS policies. Our TCRS framework introduces a controllable user simulation module that dynamically aligns with both historical personal preferences and target items preferences. 
\section{Preliminary}
\label{sec:scenario}
\begin{figure}[t]
    \centering
	\includegraphics[width=\linewidth]{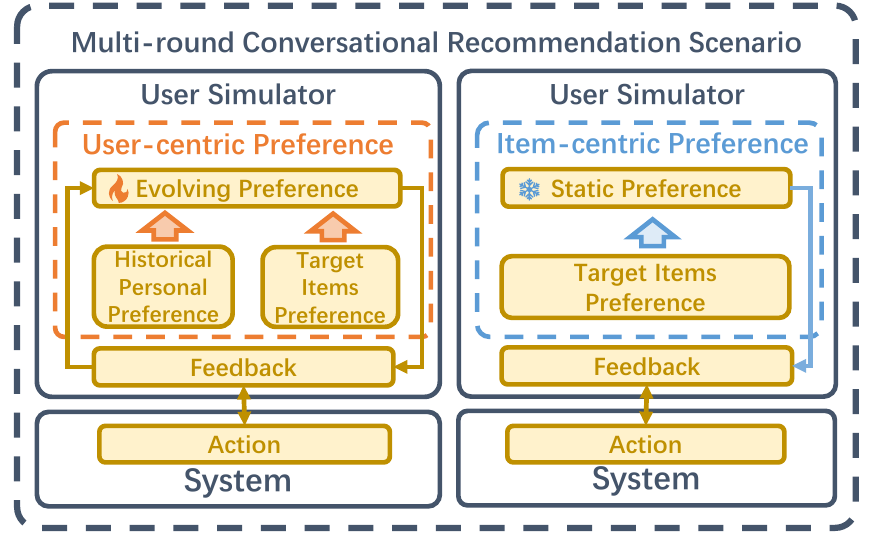}
    \caption{User-centric vs. Item-centric Preference Modeling in Multi-round Conversational Recommendation Scenario}
    \label{fig:scenario}
    \vspace{-10pt}
\end{figure}

\textbf{Multi-round Conversational Recommendation Scenario (MCR Scenario)}
The Multi-round Conversational Recommendation (MCR) scenario is a widely recognized setting in conversational recommender system (CRS) research \cite{lei20estimation,lei2020interactive}. In this scenario, the recommendation process typically involves multiple rounds of interaction between the user and the system. The conversation begins with the user expressing a specific intent, such as seeking a movie recommendation. The system can then ask the user for relevant attributes (e.g., genre, actor, director) or make recommendations. The user, in turn, provides acceptance or rejection feedback based on their preferences \cite{chu2023meta, zhang2023adaptive}.

\textbf{Limitations of Existing Item-centric Preference Modeling}: 
Existing CRS user simulators \cite{lei2020interactive,deng2021unified,zhang2022multiple, wang2023rethinking} adopt an item-centric approach to preference modeling, which assumes that user preferences are static and directly tied to the attributes of the target items. This approach fails to capture the diversity and dynamic nature of user interests. For instance, different users may favor the same item for distinct personalized reasons (e.g., one user appreciates a movie for its plot, while another values its special effects).

Furthermore, these simulators generate user feedback (accept or reject) solely based on the target items' attributes, disregarding the user's evolving preferences and their influence on personalized responses. This reductionist approach constrains the ability of CRS policies to accurately predict user reactions in real-world scenarios, where preferences are dynamic and influenced by various contextual factors.

\textbf{User-centric Preference Modeling: A Dual-Factor Approach.}
To create a more reliable evaluation environment, we propose a user-centric preference model (Figure \ref{fig:scenario}) that recognizes the dual-factor nature of user preferences. This model dynamically constructs a user's current preferences by integrating and balancing two key factors:
1) \textbf{Target Items Preferences}: The user's preferences derived from the selected attributes of the target items in the current interaction.
2) \textbf{Historical Personal Preferences}: The user's individual preferences formed from the selected attributes of items they have previously interacted with.

The user-centric preference model posits that a user's current preferences are dynamically constructed through the integration and balanced consideration of target item preferences and historical personal preferences within the specific conversation context. 

After each round of interaction with the system, the user's preferences can evolve and change, reflecting the dynamic nature of real-world user behavior. This approach aims to provide a more realistic and comprehensive simulation of user behavior in the MCR scenario, enabling a more accurate assessment of CRS model performance and their ability to adapt to dynamic user preferences.

\begin{figure*}[t]
    \centering
	\includegraphics[width=\linewidth]{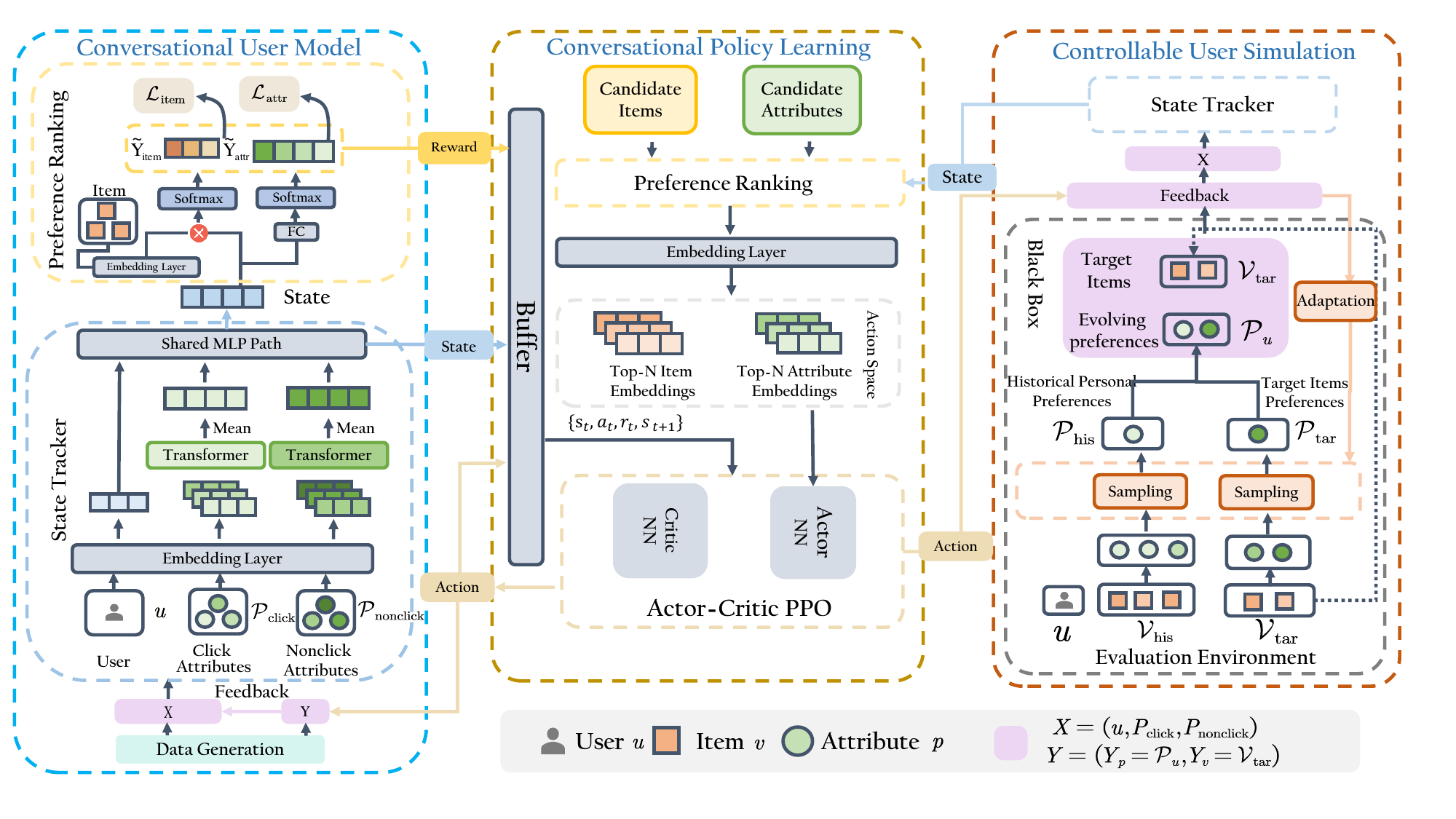}
    \caption{The Tri-Phase Offline Policy Learning-based Conversational Recommender System (TCRS) Framework. 
    Conversational User Model (CUM) is first trained on offline data to capture dynamic and personalized user preferences. The learned user model then serves as the simulated environment for Policy Learning, where the recommendation policy is optimized to maximize long-term user satisfaction. Finally, the trained policy is evaluated using an independent, controllable user simulator that can adapt to diverse user preference scenarios, enabling a comprehensive assessment of the policy's adaptability.} 
    \label{fig:method-exa}
\end{figure*}
\section{METHODOLOGY}
\subsection{Conversational User Model} 
\label{method: user_model}

The Conversational User Model adopts a multi-task learning approach, consisting of a State Tracker module and a Preference Ranking module. The key objective is to develop a comprehensive representation of user preference state that can be leveraged by the policy learning process.

\subsubsection{Data Generation for User Model}
\label{method:data_gen}
To facilitate robust training of the User Model within the TCRS framework, the Data Generation module constructs a supervised dataset \( (X, Y) \) using offline user-item pairs \( (u, v) \). This dataset is crucial for effectively simulating and learning user preferences based on historical interactions. The input feature \( X \) incorporates user \( u \) along with attributes \( P_{\text{click}} \) and \( P_{\text{nonclick}} \), which are sampled randomly from the attributes of all items previously interacted with by the user. This sampling approach promotes diversity in training data. 

Correspondingly, the label vector \( Y \) integrates attribute preferences \( \mathcal{P}_u \) derived from a combination of clicked attributes \( P_{\text{click}} \) and randomly sampled attributes of item $v$, reflecting a composite profile of user preferences. This setup ensures that the user model is trained on a varied set of potential user interactions, enhancing its ability to predict user preferences under different scenarios.

Formally, the dataset is generated as follows:
\begin{equation}
    X = (u, P_{\text{click}}, P_{\text{nonclick}})
\end{equation}
\begin{equation}
    Y = (Y_p = \mathcal{P}_u, Y_v = v)
\end{equation}
Here, \( Y_p \) and \( Y_v \) represent the attribute and item labels respectively, which are crucial for training both the State Tracker and Preference Ranking modules within the User Model.

\subsubsection{State Tracker}
State Tracker leverages a shared evolving preference encoder to capture the dynamic nature of user preferences. The encoder is designed to process both explicit and implicit preference signals from user interactions. This is achieved by separately encoding clicked and non-clicked attribute sequences, followed by integrating these with the user's embedding to form a comprehensive representation of user preferences at any given interaction state.

This encoder initially processes the learned embeddings of the clicked and non-clicked attribute sequences, denoted as \(P_{\text{click}}\) and \(P_{\text{nonclick}}\) respectively, which represent the attribute sequences. The sequences are encoded using Transformer ~\cite{vaswani2017attention} layers to capture both explicit and implicit preference signals:
\begin{equation}
    \mathbf{h}_{\text{click}} = \text{Transformer}(\text{Embed}(P_{\text{click}})),
\end{equation}
\begin{equation}
    \mathbf{h}_{\text{nonclick}} = \text{Transformer}(\text{Embed}(P_{\text{nonclick}})),
\end{equation}
where \(\text{Embed}(\cdot)\) denotes the embedding operation for attributes, and \(\mathbf{h}_{\text{click}}, \mathbf{h}_{\text{nonclick}} \in \mathbb{R}^d\) are the encoded vectors. The user's embedding, denoted by \(\mathbf{u} \in \mathbb{R}^d\), is combined with the mean pooled representations of \(\mathbf{h}_{\text{click}}\) and \(\mathbf{h}_{\text{nonclick}}\) to form the input to an MLP, producing a shared evolving preference vector \(\mathbf{s} \in \mathbb{R}^d\):
\begin{equation}
    \label{eq:state}
    \mathbf{s} = \text{MLP}(\text{Concat}(\mathbf{u}, \text{Mean}(\mathbf{h}_{\text{click}}), \text{Mean}(\mathbf{h}_{\text{nonclick}}))).
\end{equation}
This vector \(\mathbf{s}\) represents the user's current preference state, capturing both their explicit and implicit preferences.

\subsubsection{Preference Ranking}
\label{method:prefer_ranking}
The Preference Ranking module consists of two sub-tasks: item prediction and attribute prediction. The item prediction task aims to predict the user's interest in a set of candidate items based on their current preference state, while the attribute prediction task identifies the attributes that are currently preferred by the user.

\textit{Item Prediction Task.}
In the item prediction task, the objective is to predict the user's interest in a set of candidate items based on their current preference state. This is achieved by computing the dot product between the preference vector \(\mathbf{s}\) and the embeddings of candidate items, followed by a sigmoid activation to obtain probabilities:
\begin{equation}
\label{eq:item_score}
    f_{\text{item}} = \sigma(\mathbf{s}^T \cdot \text{Embed}(V_{\text{cand}})),
\end{equation}
where \(\text{Embed}(V_{\text{cand}}) \in \mathbb{R}^{m \times d}\) represents the embeddings of \(m\) candidate items, and \(f_{\text{item}} \in \mathbb{R}^m\) are the predicted probabilities of user interest in these items.

\textit{Attribute Prediction Task.}
The attribute prediction task aims to identify which attributes are currently preferred by the user, using the shared evolving preference vector \(\mathbf{s}\). Predictions are made for all attributes, with the model outputting a probability for each attribute indicating the likelihood of user interest:
\begin{equation}
    \label{eq:attribute_score}
    f_{\text{attr}} = \sigma(\mathbf{s} \cdot W_{\text{attr}} + b_{\text{attr}}).
\end{equation}
where \(f_{\text{attr}} \in \mathbb{R}^k\) represents the predicted relevance scores for \(k\) attributes, \(W_{\text{attr}} \in \mathbb{R}^{d \times k}\) is the weight matrix, and \(b_{\text{attr}} \in \mathbb{R}^k\) is the bias vector.

\textit{Joint Task Loss.}
For the item prediction task, the model optimizes a Binary Cross-Entropy (BCE) Loss, suitable for scenarios with binary outcomes. The loss for this task is given by:
\begin{equation}
\mathcal{L}_{\text{item}} = -\frac{1}{N} \sum_{v=1}^{N} \big[y_v \log(\hat{y}_v) + (1 - y_v) \log(1 - \hat{y}_v)\big],
\end{equation}
where \(N\) is the number of sample items, \(y_v\) and \(\hat{y}_v\) are the binary ground truth label and predicted probability, respectively.

For the attribute prediction task, each attribute prediction is treated as a separate binary classification problem, aggregated into a composite Binary Cross-Entropy (BCE) Loss:
\begin{equation}
\mathcal{L}_{\text{attr}} = -\frac{1}{K} \sum_{p=1}^{K} \big[y_p \log(\hat{y}_p) + (1 - y_p) \log(1 - \hat{y}_p)\big],
\end{equation}
where \(K\) denotes the number of attributes in the dataset.

The overall training objective combines the losses from both the item and attribute prediction tasks, weighted by parameters \(\lambda\) and \(1-\lambda\) to balance their contributions:
\begin{equation}
    \mathcal{L} = \lambda \mathcal{L}_{\text{item}} + (1-\lambda) \mathcal{L}_{\text{attr}}.
\end{equation}

By integrating the State Tracker and Preference Ranking modules, the Conversational User Model is designed to capture the complexity of user preferences, enabling it to serve as a robust training environment for the subsequent Conversational Policy Learning component of the TCRS framework.

\subsection{Conversational Policy Learning} 
\label{method: policy_learning}
This section will describe the implementation of the Proximal Policy Optimization (PPO) algorithm \cite{schulman2017proximal} used to derive effective decision-making strategies for the CRS. The policy network takes the user preference state and preference-based reward as input to learn actions (recommended items or queried attributes) that maximize long-term reward. This allows the CRS to learn optimal recommendation policies without the need for real-time user data.

\subsubsection{State Representation}
The state $\mathbf{s}_t$ at interaction \(t\) in our reinforcement learning (RL) framework is derived from the State Tracker of the Conversational User Model in Equation (\ref{eq:state}), capturing the current evolving user preferences. The output of the User Model's MLP encoder serves as the perceived state of the user's preference, providing a rich representation for decision-making.

\subsubsection{Action Space}
Given the potentially vast action space in CRS, efficient exploration strategies are crucial \cite{deng2021unified}. We employ the Preference Ranking of the Conversational User Model in Equation (\ref{eq:item_score}) and Equation (\ref{eq:attribute_score}) to prune the candidate item and attribute space, leveraging learned user preference predictions to narrow down choices. The action space is thus defined by selecting the top-ranked items and attributes based on Preference Ranking's probability distributions:
\begin{equation}
    \mathcal{A}_\text{space} = \mathcal{V}_{\text{top-N}}^{(t)} + \mathcal{P}_{\text{top-N}}^{(t)}.  
\end{equation}
Where \(\mathcal{V}_{\text{top-N}}^{(t)}\) and \(\mathcal{P}_{\text{top-N}}^{(t)} \) represent the top-N items in Equation (\ref{eq:item_score}) and top-N attributes in Equation (\ref{eq:attribute_score}), respectively, based on the current state \(\mathbf{s}_t\). 

\subsubsection{Reward and Termination}
\label{method:reward_done}
\textit{Reward Calculation}:
The reward at each interaction \( t \) is determined by the degree to which the actions \( \mathcal{A}_t \), taken based on the policy’s decision, align with the user's preference profile as indicated by Preference Ranking scores in Section (\ref{method:prefer_ranking}). Specifically, when the action involves querying attributes or recommending items, the reward is computed as the cumulative preference score for these actions:
\begin{equation}
    \text{reward} = \sum (\text{Preference Ranking}(s_t, \mathcal{A}_t)).
\end{equation}
This formulation ensures that the reinforcement learning algorithm is driven to maximize actions that are highly preferred by the user, reinforcing the CRS's ability to adapt to user preferences efficiently.

\textit{Termination Signal ('done' Status)}:
The 'done' status is activated either when the number of interactions reaches a predefined limit \( T \) or when any recommended item 
\( v \) matches the user's target items, as indicated by a non-empty intersection between \( \mathcal{A}_t \) and \( Y_v \):
\begin{equation}
    \text{done} = \begin{cases} 
      \text{true} & \text{if } t = T \text{ or } \mathcal{A}_t \cap Y_v \neq \emptyset  \\
      \text{false} & \text{otherwise}
   \end{cases}
\end{equation}
This setup effectively signals the completion of a CRS interaction session, either through the natural progression of dialogues or the successful recommendation of an item, thus optimizing the policy learning loop for realistic and effective user engagements.

\subsubsection{PPO Optimization}

Policy Optimization (PPO) algorithm employs two distinct Multi-Layer Perceptron (MLP) neural networks for its actor and critic components. The actor network is responsible for generating action probabilities based on the current policy, while the critic network evaluates these actions by predicting their respective state values. This separation helps to stabilize the learning updates and ensures efficient policy refinement.

The objective function of PPO, \(\mathcal{L}^{CLIP+VF+S}(\theta)\), integrates three key components:
\begin{itemize}[leftmargin=1mm]
    \item The \textbf{clipping objective} (\(\mathcal{L}^{CLIP}(\theta)\)) limits the update step, reducing the variance of policy updates and improving stability. It is defined as:
    \begin{equation}
        \mathcal{L}^{CLIP}(\theta) = \hat{\mathbb{E}}_t [\min(r_t(\theta) \hat{A}_t, \text{clip}(r_t(\theta), 1 - \epsilon, 1 + \epsilon) \hat{A}_t)].
    \end{equation}
    where \(r_t(\theta)\) is the probability ratio of the new and old policies, and \(\hat{A}_t\) is the advantage estimate at time \(t\), reflecting the relative benefit of the selected action.
    
    \item The \textbf{value function loss} (\(\mathcal{L}^{VF}(\theta)\)) measures the prediction error of the critic, encouraging accurate state value estimates.
    
    \item The \textbf{entropy bonus} (\(S[\pi_\theta](s_t)\)) promotes exploration by incentivizing the policy to maintain a diverse probability distribution over actions.
\end{itemize}

\subsection{Controllable User Simulation} 
\label{method: user_simulator}
The TCRS framework introduces the Controllable User Simulation module, which aims to provide a more realistic and comprehensive evaluation environment for CRS policies.

\subsubsection{Preference Initialization and Adaptation}
The Controllable User Simulation initializes user preferences using a parameterized approach. A key parameter, $\alpha$, dictates the balance between historical personal preferences and target items preferences, where $\alpha=0$ represents a purely item-centric preference model, and $\alpha=1$ denotes a purely personalized preference model.

During the conversational interaction, the Controllable User Simulation dynamically adjusts the personalized preference parameter $\alpha$ in response to user engagement (clicks). 
Specifically, if the user's engagement indicates a predominant interest in the attributes of the target items (more clicks than non-clicks), $\alpha$ is decreased by a preference evolution rate $\Delta\lambda$, shifting the balance towards a more pronounced consideration of the target items preference. Conversely, if the engagement shows a lesser interest (more non-clicks than clicks), $\alpha$ is increased by $\Delta\lambda$, reasserting the influence of historical personal preferences.

This adaptive mechanism enables a nuanced balance, dynamically aligning the system's recommendations with the evolving preferences of the user, as opposed to the static item-centric assumptions made by previous simulators.

\subsubsection{Preference Sampling Strategies}
\label{method: user_sim_sample}
To further enhance the realism and diversity of the user simulation, the Controllable User Simulation employs a dual-faceted approach to sampling historical personal preferences and target items preferences.

\begin{itemize}
\item \textbf{Random Sampling}: This strategy treats all attributes equally, fostering a broad exploration of item attributes. Its simplicity ensures fairness but may overlook critical attributes.
\item \textbf{Frequency-based Sampling}: By prioritizing attributes based on their historical occurrence in user interactions, this strategy emphasizes familiar preferences, potentially at the expense of new interest discovery.
\item \textbf{Inverse Frequency Sampling}: This approach highlights less common attributes, promoting the exploration of novel interests. While it aids in discovering unique user preferences, it risks overvaluing minor attributes.
\end{itemize}
These diverse sampling strategies enable the Controllable User Simulation to simulate a wide range of user behaviors. 

\subsubsection{Significance and Limited} 
The proposed rule-based, user-centric preference simulator advances CRS model evaluations by incorporating historical personal preferences and target items preferences, offering a dynamic alternative to traditional item-centric simulators. However, its rule-based design may not capture the complex dynamics of actual user preferences fully. While the simulator's controllability and versatility are beneficial, its simplicity could limit its effectiveness in mirroring true user behavior.

\section{Experiments}
To verify the effectiveness of the Tri-Phase Offline Policy Learning-based Conversational Recommender System (TCRS) framework, as well as to evaluate the experimental effects under different user-centric preference scenario settings, we use the following research questions to guide our experiments:
\begin{itemize}
    \item \textbf{RQ1.} How does the Tri-Phase Offline Policy Learning-based Conversational Recommender System (TCRS) framework's performance compare to start-of-the-art baselines?
    \item \textbf{RQ2.} What is the impact of key component of the TCRS framework on the overall performance?
    \item \textbf{RQ3.} How does the adaptability of the TCRS framework to the various user simulation scenarios evaluation?
    \item \textbf{RQ4.} What are the trends in user preference evolution during interactions with the TCRS framework?
\end{itemize}

\subsection{Dataset Description}
\begin{table}[tbp]
  \centering
  \caption{Statistics of datasets.}
    \begin{tabular}{lrrr}
    \toprule
    \textbf{Dataset} & \multicolumn{1}{l}{\textbf{LastFM}} & \multicolumn{1}{r}{\textbf{Yelp}} & \multicolumn{1}{l}{\textbf{Amazon-Book}} \\
    \midrule
     \#Users  &               1,801  &          27,675  &          30,288  \\
     \#Items  &               7,432  &          70,311  &          17,739  \\
     \#Interactions  &             76,693  &      1,368,609  &        477,304  \\
     \#Attributes  &               8,438  &               590  &              982  \\
    \midrule
     \#Avg. Attr. per Item  &               12.71  &              6.78  &              4.98  \\
     \#Avg. Item per Attr.  &               11.19  &          808.49  &            89.98  \\
    \bottomrule
    \end{tabular}%
  \label{tab:data}%
\vspace{-10pt}
\end{table}%

In this study, we analyze three datasets, each with unique features and relevance to recommendation systems, summarized in Table ~\ref{tab:data}:
\begin{itemize}
\item \textbf{LastFM\footnote{https://grouplens.org/datasets/hetrec-2011/} ~\cite{lei20estimation}:} Derived from a musician recommendation platform, the LastFM dataset applies a 5-core filter to users and items, showcasing 8,438 item-related attributes primarily from user tags. It is characterized by detailed attribute information, with a high average number of attributes per item and a lower average item count per attribute.

\item \textbf{Yelp\footnote{https://www.yelp.com/dataset/} ~\cite{lei20estimation}:} From a business recommendation context, the Yelp dataset uses a 5-core filter, featuring 590 attributes ~\cite{lei20estimation} such as city, ratings, price levels, and categories. It stands out for its extensive item coverage, with attributes linked to a higher number of items.

\item \textbf{Amazon-Book\footnote{http://jmcauley.ucsd.edu/data/amazon.} ~\cite{wang2019kgat}:} Focused on book recommendations, this dataset employs a 10-core filter on interactions, with attribute data derived from a knowledge graph. 
\end{itemize}

\subsection{Experimental Setup}
\subsubsection{User Simulator for Policy Evaluation}
\label{Exp:user_sim}
To comprehensively evaluate the performance of the CRS, we adopt the designed controllable user simulation that can dynamically model user preferences. The simulator is grounded in the multi-round conversational recommendation scenario described in Section \ref{sec:scenario}, allowing us to assess the CRS's adaptability to evolving user preferences.

The user simulator is initialized with a tuple $(u, V_{\text{his}}, V_{\text{tar}})$ , where $u$ denotes a user, $V_{\text{his}}$ represents the set of items that the user has interacted with in the training data, and $V_{\text{tar}}$ comprises the target items in the test data. Typically, \(V_{\text{tar}}\) may contain multiple items, we focus on a single target item \(v\) from the $(u, v)$ pair. This selection is intended to simplify the testing framework and provide clarity on the interaction dynamics under study.

At the core of the user simulator is the dynamic, user-centric preference model (Section \ref{method: user_simulator}), which derives user preferences from both the historical personal preferences $\mathcal{P}_{\text{his}}$ and the current target item preference $\mathcal{P}_{\text{tar}}$. When the CRS requests relevant user preferences, the simulator responds with the user's current preference states.

To fine-tune the user simulator's behavior, we introduce several key parameters:

\begin{itemize}[leftmargin=*]
\item \textbf{Initial Personalization Parameter ($\alpha$)}: This parameter controls the initial weighting between historical personal preferences $\mathcal{P}_{\text{his}}$ and target item preference $\mathcal{P}_{\text{tar}}$ in shaping the user's initial preference state. We set $\alpha=0.5$ to equally balance two factors.
\item \textbf{Preference Evolution Rate ($\Delta\lambda$)}: This parameter determines the pace at which the user's preferences evolve towards either target item preference or their historical personal preferences, based on the simulated interactions. We set $\Delta \lambda=0.1$ to facilitate a moderate rate of preference evolution.
\item \textbf{Preference Sampling Strategies}: The simulator employs frequency-based sampling for $\mathcal{P}_{\text{his}}$ to leverage salient historical personal preferences, and inverse frequency sampling for $\mathcal{P}_{\text{tar}}$ to encourage exploration of less prominent item attributes.
\item \textbf{Preference Set Size}: For consistency across simulations, we fix the user evolving preference size $|\mathcal{P}_{u}|$ to match the number of target item attributes.
\end{itemize}

By carefully configuring these parameters, the user simulator can model a diverse range of user behavior, from static to dynamic preferences (Personalization Parameter $\alpha$ and Preference Evolution parameter $\Delta\lambda$) and narrow to broad interests (Preference Sampling Strategies and Preference Set Size). This flexibility allows us to comprehensively evaluate the CRS's adaptability and performance across a variety of user scenarios, a crucial step in developing robust and user-centric personalization solutions.

\begin{table*}[htbp]
    \small
  \centering
  \caption{Statistically significant performance comparison of TCRS and baselines on Success Rate and Average Turn Metrics}
    \begin{tabular}{cccccrccccrcccc}
    \toprule
    \multirow{2}[4]{*}{\textbf{Models}} & \multicolumn{4}{c}{\textbf{LastFM}} &       & \multicolumn{4}{c}{\textbf{Yelp}} &       & \multicolumn{4}{c}{\textbf{Amazon-Book}} \\
\cmidrule{2-5}\cmidrule{7-10}\cmidrule{12-15}          & \textbf{SR@5} & \textbf{SR@10} & \textbf{SR@15} & \textbf{AT} &       & \textbf{SR@5} & \textbf{SR@10} & \textbf{SR@15} & \textbf{AT} &       & \textbf{SR@5} & \textbf{SR@10} & \textbf{SR@15} & \textbf{AT} \\
    \midrule
    EAR   & 0.219  & 0.352  & 0.405  & 11.35  &       & 0.110  & 0.194  & 0.229  & 13.08  &       & 0.133  & 0.216  & 0.251  & 12.76  \\
    SCPR  & 0.410  & 0.526  & 0.578  & 9.04  &       & 0.122  & 0.222  & 0.262  & 12.70  &       & 0.242  & 0.310  & 0.354  & 11.53  \\
    UNICORN & 0.420  & 0.556  & 0.618  & 8.78  &       & 0.144  & 0.228  & 0.288  & 12.50  &       & 0.276  & 0.370  & 0.426  & 10.88  \\
    MCMIPL & 0.414  & 0.518  & 0.574  & 9.06  &       & 0.172 & 0.258  & 0.316  & 12.23 &       & 0.274  & 0.362  & 0.414  & 10.96  \\
    VPPL & 0.412  & 0.540  & 0.620  & 8.83  &       & \textbf{0.194}  & 0.240 & 0.390  &  12.29 &       & 0.274  & 0.370  & 0.426  & 10.87  \\
    \midrule
    Abs Greedy (CUM) & \textbf{0.462 } & 0.546 & 0.660  & 8.14 &       & 0.076  & 0.144  & 0.200  & 13.41  &       & 0.301  & 0.353  & 0.381  & 10.47  \\
    Max Entropy (CUM) & 0.357  & 0.446  & 0.522  & 9.46  &       & 0.052  & 0.176  & 0.293  & 13.15  &       & 0.236  & 0.297  & 0.397  & 11.22  \\
    \midrule
    \textbf{TCRS} & 0.431  & \textbf{0.590 }  & \textbf{0.670 } & \textbf{7.58 }  &       & 0.105  & \textbf{0.326}  & \textbf{0.415 } & \textbf{11.62}  &       & \textbf{0.313 } & \textbf{0.381 } & \textbf{0.429 } & \textbf{10.26 } \\
    \bottomrule
    \end{tabular}%
  \label{tab:performance}%
\end{table*}%

\subsubsection{Baselines}
\label{Exp:baselines} 

The baseline CRS can be broadly categorized into two groups: those that rely on item-centric user simulators for policy training, and those that do not require a user simulator.

\textbf{Group 1: Simulator-Dependent Methods.} Methods in this group use item-centric user simulators, which are limited to feedback generation based solely on target item attributes. This often results in policies that perform well in simulations but not in real-world interactions due to their inability to adapt to evolving user preferences. Included methods are:
\begin{itemize}[leftmargin=*]
    \item \textbf{EAR~\cite{lei20estimation} and SCPR~\cite{lei2020interactive}:} In the Multi-round Conversational Recommendation (MCR) scenario, both methods treat recommendation and decision-making as distinct tasks, employing Factorization Machines (FM)  for recommendation and reinforcement learning for decision-making.
    \item \textbf{UNICORN~\cite{deng2021unified} and MCMIPL~\cite{zhang2022multiple}:} In MCR scenario, These are end-to-end graph-based reinforcement learning methods that model dynamic user preferences as state representations using dynamic graphs. 
    \item \textbf{VPPL~\cite{zhang2023adaptive}:} It considers a modified scenario of the MCR scenario, where the user has vague preference. It dynamically models user click and non-click behavior preferences in each round, viewing the user decision process as uncertain preference tendencies rather than absolute binary decisions (accept or reject).
\end{itemize}

\textbf{Group 2: Non-Simulator Methods.} This group includes methods that operate without a user simulator, utilizing the TCRS's Conversational User Model (CUM) as recommendation model:
\begin{itemize}[leftmargin=*]
    \item \textbf{Abs Greedy\cite{christakopoulou2016towards}:} A basic approach that continuously executes recommendations until a successful outcome. 
    \item \textbf{Max Entropy\cite{lei20estimation}:} Alternates between querying attributes with maximum information entropy and recommending items. 
\end{itemize}

\textbf{Evaluation Consistency.}
To ensure a fair and comprehensive evaluation, all baseline methods, including those that require user simulators, are assessed using the same user-centric preference simulator employed in the TCRS framework. Additionally, key policy parameters, such as the number of attribute queries or item recommendations per turn and the maximum number of interaction rounds, are kept consistent across all methods.

\subsubsection{Training Details}
\label{Exp:traing_detail}
The $(u, i)$ interaction dataset was divided into training, validation, and testing sets with a ratio of 7:1.5:1.5.

The Conversational User Model was pre-trained using offline training data. The Transformer architecture had 4 layers, with a negative sampling size of 2000 for item prediction. The batch size was set to 512, with a hidden size and embedding dimensions of 64 for users, items, and attributes. The learning rate was 0.001, using the Adam optimizer. In the multi-task training, the item prediction loss weight $\lambda$ was set to 0.8.

The Conversational Policy Learning module leveraged the pre-trained user model and offline validation data. The candidate items and attributes were pruned to the top 10 (i.e., $\text{top-N}$) based on the user model's preference rankings. The reinforcement learning training ran for $1e5$ epochs with a batch size of 2048. The actor and critic networks had a hidden size of 64, with a learning rate of $3e-4$ and a discount factor of 0.99. The entropy bonus coefficient was set to 0.01, and the clipping parameter $\epsilon$ in $L^{CLIP}$ was 0.5. A turn penalty reward of -1 was applied to encourage shorter conversations.

During the offline training and online evaluation, The policy was limited to 15 conversation turns, with a maximum of 2 attribute queries and 10 item recommendations per turn. During the evaluation, the user simulator settings are described in Section \ref{Exp:user_sim}.

\subsubsection{Evaluation Metrics}
\label{Exp:evaluation_metric}
The study utilized two key metrics to evaluate the effectiveness of the conversational recommendations \cite{lei20estimation}:

\textbf{Success Rate (SR@$T$)}: This metric quantifies the proportion of successful recommendations achieved within $T$ turns. A higher SR@$T$ indicates superior performance. 

\textbf{Average Turn (AT)}: This metric represents the average number of turns taken in a conversation. A lower AT indicates increased efficiency in the recommendation process.

\subsection{Performance Comparison (RQ1)} 
\label{exp:RQ1}


As presented in Table \ref{tab:performance}, showcases a comparative study against conventional CRS baselines. We have the following findings:
\begin{itemize}[leftmargin=1mm]
    \item The Tri-Phase Offline Policy Learning-based Conversational Recommender System (TCRS) framework demonstrates superior recommendation performance across all datasets compared to the item-centric user simulator-based baselines (EAR, SCPR, UNICORN, MCMIPL, VPPL) as well as the simulator-free baselines (Abs Greedy and Max Entropy).
    \item In smaller datasets like LastFM, Abs Greedy's exhibits competitive performance, showing the potency of Conversational User Model (CUM) component within the TCRS framework. However, it struggles in the Yelp and Amazon-Book datasets, which contain a large number of items. In comparison, Max Entropy's poor performance across all datasets highlights the importance of policy learning to capture personalized and dynamic user preferences.
    \item VPPL and TCRS outperform other CRS models as they conduct meticulous user behavior assumptions tailored to policy learning, better capturing realistic user preference modeling.
    \item  In Yelp, TCRS leverages long-term dynamic preferences more effectively despite weaker short-term performance. This shows TCRS's strength in capturing evolving user preferences long-term. The abundance of items per attribute in the Yelp dataset leads to the TCRS seeking asking more rounds of information before providing precise recommendations. 
\end{itemize}

\subsection{Evaluating Key Design in TCRS (RQ2)}
\begin{table}[htbp]
\small

  \centering
  \caption{Evaluation of the CUM component and different sampling strategies in the TCRS framework.}
  \resizebox{\columnwidth}{!}{
    \begin{tabular}{l|cc|cc|cc}
    \toprule
    \multicolumn{1}{c}{\multirow{2}[4]{*}{}} & \multicolumn{2}{c}{\textbf{LastFM}} & \multicolumn{2}{c}{\textbf{Yelp}} & \multicolumn{2}{c}{\textbf{Amazon-Book}} \\
\cmidrule{2-7}    \multicolumn{1}{c}{} & \textbf{SR@15} & \textbf{AT} & \textbf{SR@15} & \textbf{AT} & \textbf{SR@15} & \textbf{AT} \\
    \midrule
    \multicolumn{1}{c|}{\textbf{Ours}} & 0.670 & \textbf{7.58} & \textbf{0.415} & \textbf{11.62} & \textbf{0.429} & \textbf{10.26} \\
    \midrule
    (a) - w/o CUM Item & 0.682  & 8.15  & 0.257  & 12.43  & 0.349  & 11.37  \\
    (b) - w/o CUM Attr. & 0.598  & 8.89  & 0.224  & 13.12  & 0.360  & 11.27  \\
    (c) - w/o CUM Both & \textbf{0.690}  & 8.23  & 0.313  & 12.32  & 0.393  & 11.03  \\
    \midrule
    (d) - Random (His.) & 0.630  & 8.48  & 0.321  & 12.63  & 0.405  & 10.73  \\
    (e) - Random (Tar.) & 0.646  & 8.42  & 0.365  & 12.47  & 0.401  & 10.68  \\
    (f) - Frequency (Tar.) & 0.666  & 7.88  & 0.393  & 12.32  & 0.421  & 10.65  \\
    \bottomrule
    \end{tabular}%
  \label{tab:ab_study}%
  }
\end{table}%

\subsubsection{Ablation Study on CUM}

The proposed Conversational User Model (CUM) module, a key TCRS framework component, captures dynamic user preferences by jointly modeling items and attributes. 
We compare the full CUM model with two graph-based variants \cite{deng2021unified, zhang2022multiple, zhang2023adaptive}: 1) without the Item Preference Ranking module, and 2) without the Attribute Preference Ranking module.

Table \ref{tab:ab_study} (a-c) shows that removing either the item or attribute preference prediction leads to performance degradation across the Yelp and Amazon-Books datasets. This highlights the importance of CUM's joint modeling in constructing an accurate and adaptive preference action space for the downstream RL-based policy. Interestingly, the graph-based preference modeling approach (row c) performs reasonably well on the LastFM dataset. This is likely due to the denser attribute information available in LastFM, which plays to the strengths of the graph-based method. However, our full CUM model still achieves superior average turn counts, underscoring its efficiency in streamlining recommendations compared to the graph-based baseline.

Comparing rows (a-b) and (c) shows unified modeling of items and attributes outperforms decoupled prediction. This validates the interdependence of item and attribute relevance in shaping user interests.

\subsubsection{Sampling Strategies of Controllable User Simulation}
Another key TCRS component, the Controllable User Simulation module, provides a realistic evaluation by modeling the personalized and evolving nature of user preferences. A crucial aspect of this module is the sampling strategies used to generate user feedback during the conversational recommendation evaluation.

In user-centric evolving preference simulator, the aim of the proposed sampling approaches (Section ~\ref{Exp:user_sim}) is to balance exploiting known user preference and exploring item preference during conversations. Results in Table \ref{tab:ab_study} validate the merits of personalized and target item sampling for achieving this objective.
Compared to personalized frequency sampling (Ours), random personalized sampling (d) underperforms across datasets. This aligns with our motivation that frequency-based sampling better captures users' most salient historical personal preferences, while random sampling risks diluting these signals. Similarly, inverse frequency sampling on target item preference (Ours) outperforms target item preference random sampling (e). This confirms our hypothesis that highlighting less common attributes aids exploration of user preference towards target item during conversations.

Interestingly, frequency-based sampling on target item preference (f) also achieves strong performance by prioritizing prominent target attributes. This suggests both sampling logics have merits depending on the objective of generalization vs personalization.

\subsection{Adaptability Evaluation (RQ3)} 
\begin{figure}[htbp]
\begin{minipage}[t]{0.49\linewidth}
    \includegraphics[width=\linewidth]{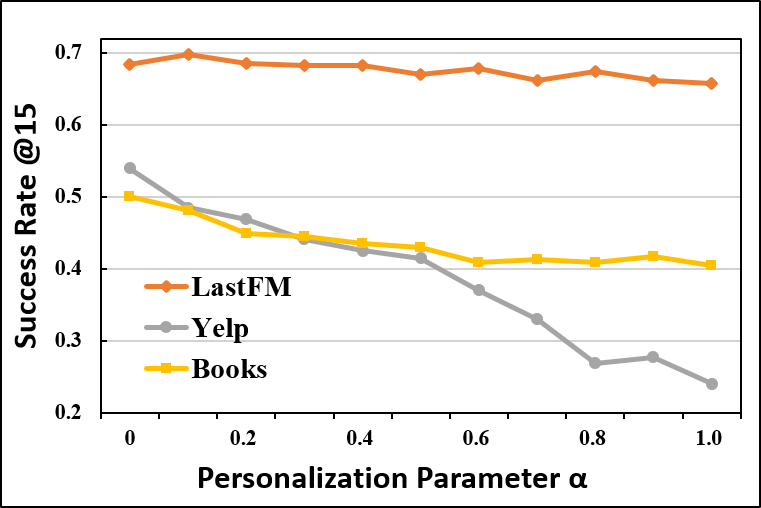}
    \label{f1}
\end{minipage}%
    \hfill%
\begin{minipage}[t]{0.49\linewidth}
    \includegraphics[width=\linewidth]{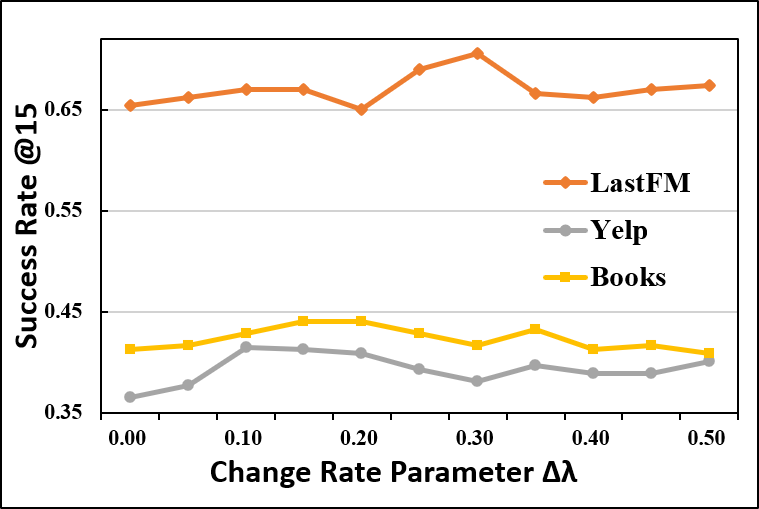} 
    \label{f2}
\end{minipage}
\vspace{-15pt}
\caption{Evaluating TCRS policies under varying personalized preference ($\alpha$) and preference evolution rate ($\Delta\lambda$) in the Controllable User Simulation (RQ3).
} 
\vspace{-10pt}
\label{fig:RQ3}
\end{figure}

A key contribution of the TCRS framework is its ability to adapt to diverse and dynamically evolving user preferences. To assess this adaptability, we leverage the Controllable User Simulation module to create varied user preference environments.

As shown in Figure \ref{fig:RQ3}, we adjust two key parameters: the personalized preference initialization ($\alpha$) and the preference evolution rate ($\Delta\lambda$). This allows us to evaluate the CRS policies' performance under different personalized and evolving user scenarios.

Increasing $\alpha$ emphasizes personalized preferences over dynamic adaptation to target item. The results show that higher $\alpha$ leads to performance declines on Yelp and Books, as policies become overly reliant on static preferences. However, LastFM exhibits stable performance, suggesting personalized preferences are sufficient for accurate recommendations in this domain. Elevating $\Delta\lambda$ introduces a more dynamic preference landscape, testing the policies' ability to adapt. The success rate remains largely stable, underscoring the TCRS framework's robustness in reconciling evolving preferences with recommendation convergence.

These results showcase the TCRS framework's effectiveness in aligning with diverse user preference characteristics, transcending the limitations of traditional item-centric simulators.

\subsection{Exploring User Preference Evolution (RQ4)}
\begin{figure}[htbp]
\begin{minipage}[t]{0.49\linewidth}
    \includegraphics[width=\linewidth]{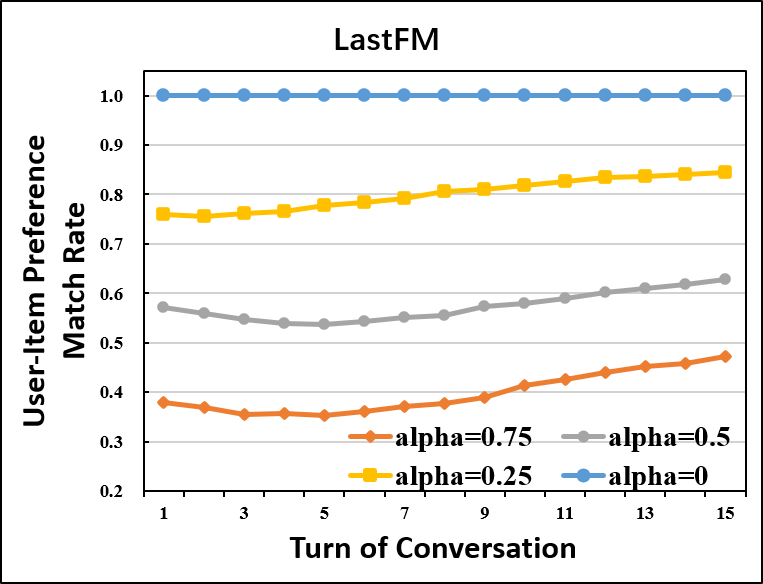}
\end{minipage}%
    \hfill%
\begin{minipage}[t]{0.49\linewidth}
    \includegraphics[width=\linewidth]{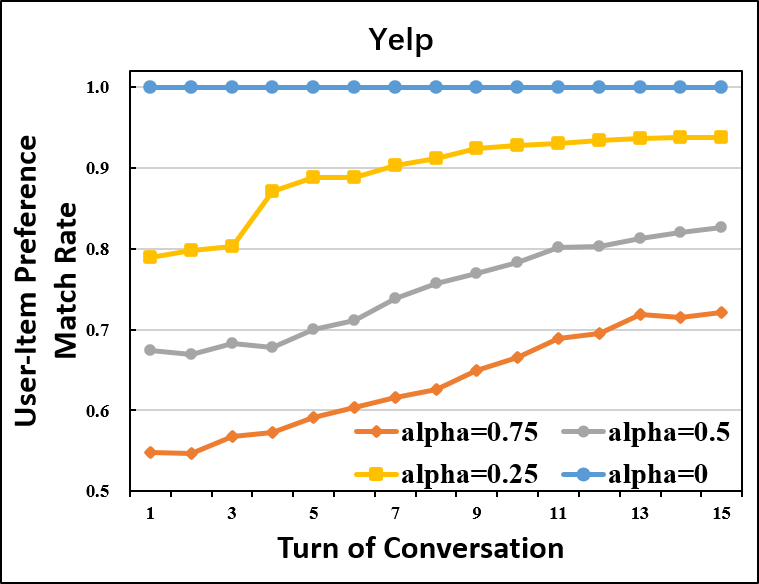} 
\end{minipage}
\caption{Examining the alignment between user preference and target item attributes (match rate) during the conversational interaction, under varying levels of personalization ($\alpha$) in the LastFM and Yelp datasets (RQ4).
} 
\vspace{-15pt}
\label{fig:RQ4}
\end{figure}

A key advantage of the TCRS framework is its ability to capture the dynamic and personalized nature of user preferences, in contrast to the static, item-centric assumptions of traditional CRS simulators. To validate this capability, we examine how the alignment between the user's evolving preferences and the attributes of the target item changes over the course of the conversational interaction.

Figure \ref{fig:RQ4} visualizes the match rate, which represents the degree of overlap between the user's current preference and the target item's attributes. This metric serves as a proxy for understanding the system's effectiveness in eliciting user interests that are well-aligned with the recommended items.

The results show that while the initial match rates differ across datasets, the overlap consistently improves as the conversation progresses. This pattern holds true regardless of the level of personalization (controlled by the $\alpha$ parameter), highlighting the TCRS framework's adaptive capacity in modeling evolving user preferences. As the system inquires about user preferences and makes recommendations, it is able to successfully align the user's interests with the target item attributes.

\section{Conclusion}

The Tri-Phase Offline Policy Learning-based Conversational Recommender System (TCRS) framework represents a significant advancement in conversational recommender systems. by addressing the critical limitations of existing approaches. By decoupling the training and evaluation phases, TCRS introduces an effective offline policy learning method that effectively mitigates the overfitting issue associated with traditional simulator-based approaches.

The explicit modeling of user preference dynamics and personalization within the Conversational User Model empowers the CRS policies to better capture and adapt to real-world user behaviors. The introduction of the Controllable User Simulation module provides a more realistic and comprehensive evaluation environment, enabling a thorough assessment of policy performance across diverse user scenarios.

The key innovations of TCRS, including the tri-phase architecture, the multi-task user preference modeling, and the user-centric simulation, pave the way for more effective and robust conversational recommender systems. These advancements allow CRS policies to better understand and cater to individual user needs, leading to enhanced user experiences and increased adoption of CRS in real-world applications.

\section{Acknowledgements}

This work is supported by the Fundamental Research Funds for the Central Universities of China (WK2100000053, PA2024GDSK0107), the National Natural Science Foundation of China (No. 62402470), and the Postdoctoral Fellowship Program of CPSF under Grant Number GZC20241643. This research is also supported by the advanced computing resources provided by the Supercomputing Center of the USTC.


\bibliographystyle{ACM-Reference-Format}
\bibliography{reference_removeURL}


\begin{thebibliography}{33}


\ifx \showCODEN    \undefined \def \showCODEN     #1{\unskip}     \fi
\ifx \showDOI      \undefined \def \showDOI       #1{#1}\fi
\ifx \showISBNx    \undefined \def \showISBNx     #1{\unskip}     \fi
\ifx \showISBNxiii \undefined \def \showISBNxiii  #1{\unskip}     \fi
\ifx \showISSN     \undefined \def \showISSN      #1{\unskip}     \fi
\ifx \showLCCN     \undefined \def \showLCCN      #1{\unskip}     \fi
\ifx \shownote     \undefined \def \shownote      #1{#1}          \fi
\ifx \showarticletitle \undefined \def \showarticletitle #1{#1}   \fi
\ifx \showURL      \undefined \def \showURL       {\relax}        \fi
\providecommand\bibfield[2]{#2}
\providecommand\bibinfo[2]{#2}
\providecommand\natexlab[1]{#1}
\providecommand\showeprint[2][]{arXiv:#2}

\bibitem[Chen et~al\mbox{.}(2019)]%
        {chen-etal-2019-towards}
\bibfield{author}{\bibinfo{person}{Qibin Chen}, \bibinfo{person}{Junyang Lin}, \bibinfo{person}{Yichang Zhang}, \bibinfo{person}{Ming Ding}, \bibinfo{person}{Yukuo Cen}, \bibinfo{person}{Hongxia Yang}, {and} \bibinfo{person}{Jie Tang}.} \bibinfo{year}{2019}\natexlab{}.
\newblock \showarticletitle{Towards Knowledge-Based Recommender Dialog System}. In \bibinfo{booktitle}{\emph{Proceedings of the 2019 Conference on Empirical Methods in Natural Language Processing and the 9th International Joint Conference on Natural Language Processing (EMNLP-IJCNLP)}}, \bibfield{editor}{\bibinfo{person}{Kentaro Inui}, \bibinfo{person}{Jing Jiang}, \bibinfo{person}{Vincent Ng}, {and} \bibinfo{person}{Xiaojun Wan}} (Eds.). \bibinfo{publisher}{Association for Computational Linguistics}, \bibinfo{address}{Hong Kong, China}, \bibinfo{pages}{1803--1813}.
\newblock


\bibitem[Christakopoulou et~al\mbox{.}(2018)]%
        {christakopoulou2018q}
\bibfield{author}{\bibinfo{person}{Konstantina Christakopoulou}, \bibinfo{person}{Alex Beutel}, \bibinfo{person}{Rui Li}, \bibinfo{person}{Sagar Jain}, {and} \bibinfo{person}{Ed~H. Chi}.} \bibinfo{year}{2018}\natexlab{}.
\newblock \showarticletitle{Q\&R: A Two-Stage Approach toward Interactive Recommendation}. In \bibinfo{booktitle}{\emph{Proceedings of the 24th ACM SIGKDD International Conference on Knowledge Discovery \& Data Mining}} (London, United Kingdom) \emph{(\bibinfo{series}{KDD '18})}. \bibinfo{publisher}{Association for Computing Machinery}, \bibinfo{address}{New York, NY, USA}, \bibinfo{pages}{139–148}.
\newblock
\showISBNx{9781450355520}


\bibitem[Christakopoulou et~al\mbox{.}(2016)]%
        {christakopoulou2016towards}
\bibfield{author}{\bibinfo{person}{Konstantina Christakopoulou}, \bibinfo{person}{Filip Radlinski}, {and} \bibinfo{person}{Katja Hofmann}.} \bibinfo{year}{2016}\natexlab{}.
\newblock \showarticletitle{Towards Conversational Recommender Systems}. In \bibinfo{booktitle}{\emph{Proceedings of the 22nd ACM SIGKDD International Conference on Knowledge Discovery and Data Mining}} (San Francisco, California, USA) \emph{(\bibinfo{series}{KDD '16})}. \bibinfo{publisher}{Association for Computing Machinery}, \bibinfo{address}{New York, NY, USA}, \bibinfo{pages}{815–824}.
\newblock
\showISBNx{9781450342322}


\bibitem[Chu et~al\mbox{.}(2023)]%
        {chu2023meta}
\bibfield{author}{\bibinfo{person}{Zhendong Chu}, \bibinfo{person}{Hongning Wang}, \bibinfo{person}{Yun Xiao}, \bibinfo{person}{Bo Long}, {and} \bibinfo{person}{Lingfei Wu}.} \bibinfo{year}{2023}\natexlab{}.
\newblock \showarticletitle{Meta Policy Learning for Cold-Start Conversational Recommendation}. In \bibinfo{booktitle}{\emph{Proceedings of the Sixteenth ACM International Conference on Web Search and Data Mining}} (Singapore, Singapore) \emph{(\bibinfo{series}{WSDM '23})}. \bibinfo{publisher}{Association for Computing Machinery}, \bibinfo{address}{New York, NY, USA}, \bibinfo{pages}{222–230}.
\newblock
\showISBNx{9781450394079}


\bibitem[Dao et~al\mbox{.}(2024)]%
        {dao2024broadening}
\bibfield{author}{\bibinfo{person}{Huy Dao}, \bibinfo{person}{Yang Deng}, \bibinfo{person}{Dung~D. Le}, {and} \bibinfo{person}{Lizi Liao}.} \bibinfo{year}{2024}\natexlab{}.
\newblock \showarticletitle{Broadening the View: Demonstration-augmented Prompt Learning for Conversational Recommendation}. In \bibinfo{booktitle}{\emph{Proceedings of the 47th International ACM SIGIR Conference on Research and Development in Information Retrieval}} (Washington DC, USA) \emph{(\bibinfo{series}{SIGIR '24})}. \bibinfo{publisher}{Association for Computing Machinery}, \bibinfo{address}{New York, NY, USA}, \bibinfo{pages}{785–795}.
\newblock
\showISBNx{9798400704314}


\bibitem[Deffayet et~al\mbox{.}(2023)]%
        {deffayet2023offline}
\bibfield{author}{\bibinfo{person}{Romain Deffayet}, \bibinfo{person}{Thibaut Thonet}, \bibinfo{person}{Jean-Michel Renders}, {and} \bibinfo{person}{Maarten de Rijke}.} \bibinfo{year}{2023}\natexlab{}.
\newblock \showarticletitle{Offline Evaluation for Reinforcement Learning-Based Recommendation: A Critical Issue and Some Alternatives}.
\newblock \bibinfo{journal}{\emph{SIGIR Forum}} \bibinfo{volume}{56}, \bibinfo{number}{2}, Article \bibinfo{articleno}{3} (\bibinfo{date}{jan} \bibinfo{year}{2023}), \bibinfo{numpages}{14}~pages.
\newblock
\showISSN{0163-5840}


\bibitem[Deng et~al\mbox{.}(2021)]%
        {deng2021unified}
\bibfield{author}{\bibinfo{person}{Yang Deng}, \bibinfo{person}{Yaliang Li}, \bibinfo{person}{Fei Sun}, \bibinfo{person}{Bolin Ding}, {and} \bibinfo{person}{Wai Lam}.} \bibinfo{year}{2021}\natexlab{}.
\newblock \showarticletitle{Unified Conversational Recommendation Policy Learning via Graph-based Reinforcement Learning}. In \bibinfo{booktitle}{\emph{Proceedings of the 44th International ACM SIGIR Conference on Research and Development in Information Retrieval}} (Virtual Event, Canada) \emph{(\bibinfo{series}{SIGIR '21})}. \bibinfo{publisher}{Association for Computing Machinery}, \bibinfo{address}{New York, NY, USA}, \bibinfo{pages}{1431–1441}.
\newblock
\showISBNx{9781450380379}


\bibitem[Deng et~al\mbox{.}(2024)]%
        {deng2024towards}
\bibfield{author}{\bibinfo{person}{Yang Deng}, \bibinfo{person}{Lizi Liao}, \bibinfo{person}{Zhonghua Zheng}, \bibinfo{person}{Grace~Hui Yang}, {and} \bibinfo{person}{Tat-Seng Chua}.} \bibinfo{year}{2024}\natexlab{}.
\newblock \showarticletitle{Towards Human-centered Proactive Conversational Agents}. In \bibinfo{booktitle}{\emph{Proceedings of the 47th International ACM SIGIR Conference on Research and Development in Information Retrieval}} (Washington DC, USA) \emph{(\bibinfo{series}{SIGIR '24})}. \bibinfo{publisher}{Association for Computing Machinery}, \bibinfo{address}{New York, NY, USA}, \bibinfo{pages}{807–818}.
\newblock
\showISBNx{9798400704314}


\bibitem[Friedman et~al\mbox{.}(2023)]%
        {friedman2023leveraging}
\bibfield{author}{\bibinfo{person}{Luke Friedman}, \bibinfo{person}{Sameer Ahuja}, \bibinfo{person}{David Allen}, \bibinfo{person}{Zhenning Tan}, \bibinfo{person}{Hakim Sidahmed}, \bibinfo{person}{Changbo Long}, \bibinfo{person}{Jun Xie}, \bibinfo{person}{Gabriel Schubiner}, \bibinfo{person}{Ajay Patel}, \bibinfo{person}{Harsh Lara}, {et~al\mbox{.}}} \bibinfo{year}{2023}\natexlab{}.
\newblock \showarticletitle{Leveraging large language models in conversational recommender systems}.
\newblock \bibinfo{journal}{\emph{arXiv preprint arXiv:2305.07961}} (\bibinfo{year}{2023}).
\newblock


\bibitem[Gao et~al\mbox{.}(2023a)]%
        {gao2023alleviating}
\bibfield{author}{\bibinfo{person}{Chongming Gao}, \bibinfo{person}{Kexin Huang}, \bibinfo{person}{Jiawei Chen}, \bibinfo{person}{Yuan Zhang}, \bibinfo{person}{Biao Li}, \bibinfo{person}{Peng Jiang}, \bibinfo{person}{Shiqi Wang}, \bibinfo{person}{Zhong Zhang}, {and} \bibinfo{person}{Xiangnan He}.} \bibinfo{year}{2023}\natexlab{a}.
\newblock \showarticletitle{Alleviating Matthew Effect of Offline Reinforcement Learning in Interactive Recommendation}. In \bibinfo{booktitle}{\emph{Proceedings of the 46th International ACM SIGIR Conference on Research and Development in Information Retrieval}} (Taipei, Taiwan) \emph{(\bibinfo{series}{SIGIR '23})}. \bibinfo{numpages}{11}~pages.
\newblock


\bibitem[Gao et~al\mbox{.}(2021)]%
        {gao2021advances}
\bibfield{author}{\bibinfo{person}{Chongming Gao}, \bibinfo{person}{Wenqiang Lei}, \bibinfo{person}{Xiangnan He}, \bibinfo{person}{Maarten {de Rijke}}, {and} \bibinfo{person}{Tat-Seng Chua}.} \bibinfo{year}{2021}\natexlab{}.
\newblock \showarticletitle{Advances and Challenges in Conversational Recommender Systems: A Survey}.
\newblock \bibinfo{journal}{\emph{AI Open}}  \bibinfo{volume}{2} (\bibinfo{year}{2021}), \bibinfo{pages}{100--126}.
\newblock
\showISSN{2666-6510}


\bibitem[Gao et~al\mbox{.}(2022)]%
        {gao2022kuairec}
\bibfield{author}{\bibinfo{person}{Chongming Gao}, \bibinfo{person}{Shijun Li}, \bibinfo{person}{Wenqiang Lei}, \bibinfo{person}{Jiawei Chen}, \bibinfo{person}{Biao Li}, \bibinfo{person}{Peng Jiang}, \bibinfo{person}{Xiangnan He}, \bibinfo{person}{Jiaxin Mao}, {and} \bibinfo{person}{Tat-Seng Chua}.} \bibinfo{year}{2022}\natexlab{}.
\newblock \showarticletitle{KuaiRec: A Fully-observed Dataset and Insights for Evaluating Recommender Systems}. In \bibinfo{booktitle}{\emph{Proceedings of the 31st ACM International Conference on Information and Knowledge Management}} (Atlanta, GA, USA) \emph{(\bibinfo{series}{CIKM '22})}. \bibinfo{numpages}{11}~pages.
\newblock


\bibitem[Gao et~al\mbox{.}(2023b)]%
        {gao2023cirs}
\bibfield{author}{\bibinfo{person}{Chongming Gao}, \bibinfo{person}{Shiqi Wang}, \bibinfo{person}{Shijun Li}, \bibinfo{person}{Jiawei Chen}, \bibinfo{person}{Xiangnan He}, \bibinfo{person}{Wenqiang Lei}, \bibinfo{person}{Biao Li}, \bibinfo{person}{Yuan Zhang}, {and} \bibinfo{person}{Peng Jiang}.} \bibinfo{year}{2023}\natexlab{b}.
\newblock \showarticletitle{CIRS: Bursting Filter Bubbles by Counterfactual Interactive Recommender System}.
\newblock \bibinfo{journal}{\emph{ACM Trans. Inf. Syst.}} \bibinfo{volume}{42}, \bibinfo{number}{1}, Article \bibinfo{articleno}{14} (\bibinfo{date}{aug} \bibinfo{year}{2023}), \bibinfo{numpages}{27}~pages.
\newblock
\showISSN{1046-8188}


\bibitem[Lei et~al\mbox{.}(2020a)]%
        {lei20estimation}
\bibfield{author}{\bibinfo{person}{Wenqiang Lei}, \bibinfo{person}{Xiangnan He}, \bibinfo{person}{Yisong Miao}, \bibinfo{person}{Qingyun Wu}, \bibinfo{person}{Richang Hong}, \bibinfo{person}{Min-Yen Kan}, {and} \bibinfo{person}{Tat-Seng Chua}.} \bibinfo{year}{2020}\natexlab{a}.
\newblock \showarticletitle{Estimation-Action-Reflection: Towards Deep Interaction Between Conversational and Recommender Systems}. In \bibinfo{booktitle}{\emph{Proceedings of the 13th International Conference on Web Search and Data Mining}} (Houston, TX, USA) \emph{(\bibinfo{series}{WSDM '20})}. \bibinfo{publisher}{Association for Computing Machinery}, \bibinfo{address}{New York, NY, USA}, \bibinfo{pages}{304–312}.
\newblock
\showISBNx{9781450368223}


\bibitem[Lei et~al\mbox{.}(2020b)]%
        {lei2020interactive}
\bibfield{author}{\bibinfo{person}{Wenqiang Lei}, \bibinfo{person}{Gangyi Zhang}, \bibinfo{person}{Xiangnan He}, \bibinfo{person}{Yisong Miao}, \bibinfo{person}{Xiang Wang}, \bibinfo{person}{Liang Chen}, {and} \bibinfo{person}{Tat-Seng Chua}.} \bibinfo{year}{2020}\natexlab{b}.
\newblock \showarticletitle{Interactive Path Reasoning on Graph for Conversational Recommendation}. In \bibinfo{booktitle}{\emph{Proceedings of the 26th ACM SIGKDD International Conference on Knowledge Discovery \& Data Mining}} \emph{(\bibinfo{series}{KDD '20})}. \bibinfo{pages}{2073–2083}.
\newblock
\showISBNx{9781450379984}


\bibitem[Li et~al\mbox{.}(2018)]%
        {nips18/DeepConv}
\bibfield{author}{\bibinfo{person}{Raymond Li}, \bibinfo{person}{Samira Kahou}, \bibinfo{person}{Hannes Schulz}, \bibinfo{person}{Vincent Michalski}, \bibinfo{person}{Laurent Charlin}, {and} \bibinfo{person}{Chris Pal}.} \bibinfo{year}{2018}\natexlab{}.
\newblock \showarticletitle{Towards deep conversational recommendations}. In \bibinfo{booktitle}{\emph{Proceedings of the 32nd International Conference on Neural Information Processing Systems}} (Montr\'{e}al, Canada) \emph{(\bibinfo{series}{NIPS'18})}. \bibinfo{publisher}{Curran Associates Inc.}, \bibinfo{address}{Red Hook, NY, USA}, \bibinfo{pages}{9748–9758}.
\newblock


\bibitem[Li et~al\mbox{.}(2021)]%
        {li2020seamlessly}
\bibfield{author}{\bibinfo{person}{Shijun Li}, \bibinfo{person}{Wenqiang Lei}, \bibinfo{person}{Qingyun Wu}, \bibinfo{person}{Xiangnan He}, \bibinfo{person}{Peng Jiang}, {and} \bibinfo{person}{Tat-Seng Chua}.} \bibinfo{year}{2021}\natexlab{}.
\newblock \showarticletitle{Seamlessly Unifying Attributes and Items: Conversational Recommendation for Cold-Start Users}.
\newblock \bibinfo{journal}{\emph{ACM Trans. Inf. Syst.}} \bibinfo{volume}{39}, \bibinfo{number}{4}, Article \bibinfo{articleno}{40} (\bibinfo{date}{aug} \bibinfo{year}{2021}), \bibinfo{numpages}{29}~pages.
\newblock
\showISSN{1046-8188}


\bibitem[Liao et~al\mbox{.}(2022)]%
        {liao2020topic}
\bibfield{author}{\bibinfo{person}{Lizi Liao}, \bibinfo{person}{Ryuichi Takanobu}, \bibinfo{person}{Yunshan Ma}, \bibinfo{person}{Xun Yang}, \bibinfo{person}{Minlie Huang}, {and} \bibinfo{person}{Tat-Seng Chua}.} \bibinfo{year}{2022}\natexlab{}.
\newblock \showarticletitle{Topic-Guided Conversational Recommender in Multiple Domains}.
\newblock \bibinfo{journal}{\emph{IEEE Transactions on Knowledge and Data Engineering}} \bibinfo{volume}{34}, \bibinfo{number}{5} (\bibinfo{year}{2022}), \bibinfo{pages}{2485--2496}.
\newblock
\urldef\tempurl%
\url{https://doi.org/10.1109/TKDE.2020.3008563}
\showDOI{\tempurl}


\bibitem[Schulman et~al\mbox{.}(2017)]%
        {schulman2017proximal}
\bibfield{author}{\bibinfo{person}{John Schulman}, \bibinfo{person}{Filip Wolski}, \bibinfo{person}{Prafulla Dhariwal}, \bibinfo{person}{Alec Radford}, {and} \bibinfo{person}{Oleg Klimov}.} \bibinfo{year}{2017}\natexlab{}.
\newblock \showarticletitle{Proximal policy optimization algorithms}.
\newblock \bibinfo{journal}{\emph{arXiv preprint arXiv:1707.06347}} (\bibinfo{year}{2017}).
\newblock


\bibitem[Shi et~al\mbox{.}(2024)]%
        {billp2024}
\bibfield{author}{\bibinfo{person}{Wentao Shi}, \bibinfo{person}{Xiangnan He}, \bibinfo{person}{Yang Zhang}, \bibinfo{person}{Chongming Gao}, \bibinfo{person}{Xinyue Li}, \bibinfo{person}{Jizhi Zhang}, \bibinfo{person}{Qifan Wang}, {and} \bibinfo{person}{Fuli Feng}.} \bibinfo{year}{2024}\natexlab{}.
\newblock \showarticletitle{Large Language Models are Learnable Planners for Long-Term Recommendation}. In \bibinfo{booktitle}{\emph{Proceedings of the 47th International ACM SIGIR Conference on Research and Development in Information Retrieval}} (Washington DC, USA) \emph{(\bibinfo{series}{SIGIR '24})}. \bibinfo{pages}{1893–1903}.
\newblock
\showISBNx{9798400704314}


\bibitem[Sun and Zhang(2018)]%
        {Sun:2018:CRS:3209978.3210002}
\bibfield{author}{\bibinfo{person}{Yueming Sun} {and} \bibinfo{person}{Yi Zhang}.} \bibinfo{year}{2018}\natexlab{}.
\newblock \showarticletitle{Conversational Recommender System}. In \bibinfo{booktitle}{\emph{The 41st International ACM SIGIR Conference on Research \& Development in Information Retrieval}} (Ann Arbor, MI, USA) \emph{(\bibinfo{series}{SIGIR '18})}. \bibinfo{publisher}{Association for Computing Machinery}, \bibinfo{address}{New York, NY, USA}, \bibinfo{pages}{235–244}.
\newblock
\showISBNx{9781450356572}


\bibitem[Vaswani et~al\mbox{.}(2017)]%
        {vaswani2017attention}
\bibfield{author}{\bibinfo{person}{Ashish Vaswani}, \bibinfo{person}{Noam Shazeer}, \bibinfo{person}{Niki Parmar}, \bibinfo{person}{Jakob Uszkoreit}, \bibinfo{person}{Llion Jones}, \bibinfo{person}{Aidan~N. Gomez}, \bibinfo{person}{\L{}ukasz Kaiser}, {and} \bibinfo{person}{Illia Polosukhin}.} \bibinfo{year}{2017}\natexlab{}.
\newblock \showarticletitle{Attention is all you need}. In \bibinfo{booktitle}{\emph{Proceedings of the 31st International Conference on Neural Information Processing Systems}} (Long Beach, California, USA) \emph{(\bibinfo{series}{NIPS'17})}. \bibinfo{publisher}{Curran Associates Inc.}, \bibinfo{address}{Red Hook, NY, USA}, \bibinfo{pages}{6000–6010}.
\newblock
\showISBNx{9781510860964}


\bibitem[Wang et~al\mbox{.}(2019)]%
        {wang2019kgat}
\bibfield{author}{\bibinfo{person}{Xiang Wang}, \bibinfo{person}{Xiangnan He}, \bibinfo{person}{Yixin Cao}, \bibinfo{person}{Meng Liu}, {and} \bibinfo{person}{Tat-Seng Chua}.} \bibinfo{year}{2019}\natexlab{}.
\newblock \showarticletitle{KGAT: Knowledge Graph Attention Network for Recommendation}. In \bibinfo{booktitle}{\emph{Proceedings of the 25th ACM SIGKDD International Conference on Knowledge Discovery \& Data Mining}} (Anchorage, AK, USA) \emph{(\bibinfo{series}{KDD '19})}. \bibinfo{publisher}{Association for Computing Machinery}, \bibinfo{address}{New York, NY, USA}, \bibinfo{pages}{950–958}.
\newblock
\showISBNx{9781450362016}


\bibitem[Wang et~al\mbox{.}(2023)]%
        {wang2023rethinking}
\bibfield{author}{\bibinfo{person}{Xiaolei Wang}, \bibinfo{person}{Xinyu Tang}, \bibinfo{person}{Xin Zhao}, \bibinfo{person}{Jingyuan Wang}, {and} \bibinfo{person}{Ji-Rong Wen}.} \bibinfo{year}{2023}\natexlab{}.
\newblock \showarticletitle{Rethinking the Evaluation for Conversational Recommendation in the Era of Large Language Models}. In \bibinfo{booktitle}{\emph{Proceedings of the 2023 Conference on Empirical Methods in Natural Language Processing}}, \bibfield{editor}{\bibinfo{person}{Houda Bouamor}, \bibinfo{person}{Juan Pino}, {and} \bibinfo{person}{Kalika Bali}} (Eds.). \bibinfo{publisher}{Association for Computational Linguistics}, \bibinfo{address}{Singapore}, \bibinfo{pages}{10052--10065}.
\newblock


\bibitem[Wu et~al\mbox{.}(2023)]%
        {wu2022state}
\bibfield{author}{\bibinfo{person}{Yuxia Wu}, \bibinfo{person}{Lizi Liao}, \bibinfo{person}{Gangyi Zhang}, \bibinfo{person}{Wenqiang Lei}, \bibinfo{person}{Guoshuai Zhao}, \bibinfo{person}{Xueming Qian}, {and} \bibinfo{person}{Tat-Seng Chua}.} \bibinfo{year}{2023}\natexlab{}.
\newblock \showarticletitle{State Graph Reasoning for Multimodal Conversational Recommendation}.
\newblock \bibinfo{journal}{\emph{IEEE Transactions on Multimedia}}  \bibinfo{volume}{25} (\bibinfo{year}{2023}), \bibinfo{pages}{3113--3124}.
\newblock
\urldef\tempurl%
\url{https://doi.org/10.1109/TMM.2022.3155900}
\showDOI{\tempurl}


\bibitem[Xin et~al\mbox{.}(2022)]%
        {xin2022rethinking}
\bibfield{author}{\bibinfo{person}{Xin Xin}, \bibinfo{person}{Tiago Pimentel}, \bibinfo{person}{Alexandros Karatzoglou}, \bibinfo{person}{Pengjie Ren}, \bibinfo{person}{Konstantina Christakopoulou}, {and} \bibinfo{person}{Zhaochun Ren}.} \bibinfo{year}{2022}\natexlab{}.
\newblock \showarticletitle{Rethinking Reinforcement Learning for Recommendation: A Prompt Perspective}. In \bibinfo{booktitle}{\emph{Proceedings of the 45th International ACM SIGIR Conference on Research and Development in Information Retrieval}} (Madrid, Spain) \emph{(\bibinfo{series}{SIGIR '22})}. \bibinfo{publisher}{Association for Computing Machinery}, \bibinfo{address}{New York, NY, USA}, \bibinfo{pages}{1347–1357}.
\newblock
\showISBNx{9781450387323}


\bibitem[Yang et~al\mbox{.}(2024)]%
        {yang2024generate}
\bibfield{author}{\bibinfo{person}{Zhengyi Yang}, \bibinfo{person}{Jiancan Wu}, \bibinfo{person}{Zhicai Wang}, \bibinfo{person}{Xiang Wang}, \bibinfo{person}{Yancheng Yuan}, {and} \bibinfo{person}{Xiangnan He}.} \bibinfo{year}{2024}\natexlab{}.
\newblock \showarticletitle{Generate what you prefer: Reshaping sequential recommendation via guided diffusion}.
\newblock \bibinfo{journal}{\emph{Advances in Neural Information Processing Systems}}  \bibinfo{volume}{36} (\bibinfo{year}{2024}).
\newblock


\bibitem[Yu et~al\mbox{.}(2024)]%
        {EasyRL4Rec}
\bibfield{author}{\bibinfo{person}{Yuanqing Yu}, \bibinfo{person}{Chongming Gao}, \bibinfo{person}{Jiawei Chen}, \bibinfo{person}{Heng Tang}, \bibinfo{person}{Yuefeng Sun}, \bibinfo{person}{Qian Chen}, \bibinfo{person}{Weizhi Ma}, {and} \bibinfo{person}{Min Zhang}.} \bibinfo{year}{2024}\natexlab{}.
\newblock \showarticletitle{EasyRL4Rec: An Easy-to-use Library for Reinforcement Learning Based Recommender Systems}. In \bibinfo{booktitle}{\emph{Proceedings of the 47th International ACM SIGIR Conference on Research and Development in Information Retrieval}} \emph{(\bibinfo{series}{SIGIR '24})}. \bibinfo{pages}{977–987}.
\newblock
\showISBNx{9798400704314}


\bibitem[Zhang(2023)]%
        {zhang2023user}
\bibfield{author}{\bibinfo{person}{Gangyi Zhang}.} \bibinfo{year}{2023}\natexlab{}.
\newblock \showarticletitle{User-Centric Conversational Recommendation: Adapting the Need of User with Large Language Models}. In \bibinfo{booktitle}{\emph{Proceedings of the 17th ACM Conference on Recommender Systems}} (Singapore, Singapore) \emph{(\bibinfo{series}{RecSys '23})}. \bibinfo{publisher}{Association for Computing Machinery}, \bibinfo{address}{New York, NY, USA}, \bibinfo{pages}{1349–1354}.
\newblock
\showISBNx{9798400702419}


\bibitem[Zhang et~al\mbox{.}(2023)]%
        {zhang2023adaptive}
\bibfield{author}{\bibinfo{person}{Gangyi Zhang}, \bibinfo{person}{Chongming Gao}, \bibinfo{person}{Wenqiang Lei}, \bibinfo{person}{Xiaojie Guo}, \bibinfo{person}{Shijun Li}, \bibinfo{person}{Hongshen Chen}, \bibinfo{person}{Zhuozhi Ding}, \bibinfo{person}{Sulong Xu}, {and} \bibinfo{person}{Lingfei Wu}.} \bibinfo{year}{2023}\natexlab{}.
\newblock \showarticletitle{Adaptive Vague Preference Policy Learning for Multi-round Conversational Recommendation}.
\newblock \bibinfo{journal}{\emph{arXiv preprint arXiv:2306.04487}} (\bibinfo{year}{2023}).
\newblock


\bibitem[Zhang and Balog(2020)]%
        {zhang2020evaluating}
\bibfield{author}{\bibinfo{person}{Shuo Zhang} {and} \bibinfo{person}{Krisztian Balog}.} \bibinfo{year}{2020}\natexlab{}.
\newblock \showarticletitle{Evaluating Conversational Recommender Systems via User Simulation}. In \bibinfo{booktitle}{\emph{Proceedings of the 26th ACM SIGKDD International Conference on Knowledge Discovery \& Data Mining}} (Virtual Event, CA, USA) \emph{(\bibinfo{series}{KDD '20})}. \bibinfo{publisher}{Association for Computing Machinery}, \bibinfo{address}{New York, NY, USA}, \bibinfo{pages}{1512–1520}.
\newblock
\showISBNx{9781450379984}


\bibitem[Zhang et~al\mbox{.}(2022)]%
        {zhang2022multiple}
\bibfield{author}{\bibinfo{person}{Yiming Zhang}, \bibinfo{person}{Lingfei Wu}, \bibinfo{person}{Qi Shen}, \bibinfo{person}{Yitong Pang}, \bibinfo{person}{Zhihua Wei}, \bibinfo{person}{Fangli Xu}, \bibinfo{person}{Bo Long}, {and} \bibinfo{person}{Jian Pei}.} \bibinfo{year}{2022}\natexlab{}.
\newblock \showarticletitle{Multiple Choice Questions based Multi-Interest Policy Learning for Conversational Recommendation}. In \bibinfo{booktitle}{\emph{Proceedings of the ACM Web Conference 2022}} (Virtual Event, Lyon, France) \emph{(\bibinfo{series}{WWW '22})}. \bibinfo{publisher}{Association for Computing Machinery}, \bibinfo{address}{New York, NY, USA}, \bibinfo{pages}{2153–2162}.
\newblock
\showISBNx{9781450390965}


\bibitem[Zhou et~al\mbox{.}(2020)]%
        {zhou2020improving}
\bibfield{author}{\bibinfo{person}{Kun Zhou}, \bibinfo{person}{Wayne~Xin Zhao}, \bibinfo{person}{Shuqing Bian}, \bibinfo{person}{Yuanhang Zhou}, \bibinfo{person}{Ji-Rong Wen}, {and} \bibinfo{person}{Jingsong Yu}.} \bibinfo{year}{2020}\natexlab{}.
\newblock \showarticletitle{Improving Conversational Recommender Systems via Knowledge Graph based Semantic Fusion}. In \bibinfo{booktitle}{\emph{Proceedings of the 26th ACM SIGKDD International Conference on Knowledge Discovery \& Data Mining}} (Virtual Event, CA, USA) \emph{(\bibinfo{series}{KDD '20})}. \bibinfo{publisher}{Association for Computing Machinery}, \bibinfo{address}{New York, NY, USA}, \bibinfo{pages}{1006–1014}.
\newblock
\showISBNx{9781450379984}


\end{thebibliography}


\end{document}